\documentclass[12pt,a4paper]{article}

\usepackage{fancybox}       
\usepackage[english]{babel}
\usepackage{amsmath}
\usepackage{amssymb}
\usepackage{amsfonts}
\usepackage{fullpage}                    
\usepackage{graphicx,subfigure}
\usepackage{graphics}
\usepackage{color}
\usepackage{soul}
\usepackage{sidecap}
\usepackage{hyperref}

\usepackage{colortbl}
\usepackage{multirow}
\usepackage[table,xcdraw]{xcolor}
\usepackage{authblk}

\usepackage{enumerate}

\usepackage{natbib}

\usepackage{soul}
\usepackage{sidecap}
\usepackage[small,bf]{caption}
\usepackage{setspace}
\usepackage{hyperref}
\usepackage{ulem}
\newcommand{\bc}{\begin{center}}
\newcommand{\ec}{\end{center}}
\newcommand{\be}{\begin{equation}}
\newcommand{\ee}{\end{equation}}
\newcommand{\ba}{\begin{eqnarray*}}
\newcommand{\ea}{\end{eqnarray*}}
\newcommand{\by}{\begin{array}{cc}}
\newcommand{\ey}{\end{array}}
\newcommand{\bi}{\begin{itemize}}
\newcommand{\ei}{\end{itemize}}
\newcommand{\ben}{\begin{enumerate}}
\newcommand{\een}{\end{enumerate}}

\newcommand{\expect}[1]{\langle #1 \rangle}

\def\slfrac#1#2{{\mathord{\mathchoice   
        {\kern.1em\raise.5ex\hbox{$\scriptstyle#1$}\kern-.1em
        /\kern-.15em\lower.25ex\hbox{$\scriptstyle#2$}}
        {\kern.1em\raise.5ex\hbox{$\scriptstyle#1$}\kern-.1em
        /\kern-.15em\lower.25ex\hbox{$\scriptstyle#2$}}
        {\kern.1em\raise.4ex\hbox{$\scriptscriptstyle#1$}\kern-.1em
        /\kern-.14em\lower.25ex\hbox{$\scriptscriptstyle#2$}}
        {\kern.1em\raise.2ex\hbox{$\scriptscriptstyle#1$}\kern-.1em
        /\kern-.1em\lower.25ex\hbox{$\scriptscriptstyle#2$}}}}}

\begin{document}

\newcommand\sa[1]{\textcolor{blue}{#1}} 
\newcommand\gz[1]{\textcolor{red}{#1}} 
\newcommand\jb[1]{\textcolor{magenta}{#1}} 

\newcommand\blue[1]{\textcolor{blue}{#1}}
\newcommand\red[1]{\textcolor{red}{#1}}
\newcommand\green[1]{\textcolor{green}{#1}}
\newcommand\magenta[1]{\textcolor{magenta}{#1}}
\newcommand\figtitle[1]{\textit{\textbf{#1}}}
\parindent=1cm

\title{Perceptual reversals in binocular rivalry: \\ improved detection from OKN}

\author[1,2]{Stepan Aleshin}
\author[3,4]{Gerg\H{o} Ziman}
\author[3,4]{Ilona Kov\'{a}cs}
\author[1,2]{Jochen Braun}

\affil[1]{Institute of Biology, Otto-von-Guericke University, Leipzigerstr.\,44, 39120 Magdeburg, Germany}
\affil[2]{Center for Behavioral Brain Sciences, Leipzigerstr.,44, 39120 Magdeburg, Germany}
\affil[3]{Department of General Psychology, Institute of Psychology, P\'{a}zm\'{a}ny P\'{e}ter Catholic University, 1088 Budapest, Hungary}
\affil[4]{MTA-PPKE Adolescent Development Research Group, 1088 Budapest, Hungary}

\renewcommand\Authfont{\sffamily \bfseries \itshape \raggedright}
\renewcommand\Affilfont{\sffamily  \mdseries \itshape\small \raggedright}

\date{}  

\maketitle

\pagebreak

\begin{abstract}

When binocular rivalry is induced by opponent motion displays, perceptual reversals are often associated with changed oculomotor behaviour \citep{frassle2014binocular,fujiwara2017optokinetic}.   Specifically, the direction of smooth pursuit phases in optokinetic nystagmus (OKN) typically corresponds to the direction of motion that dominates perceptual appearance at any given time.  Here we report an improved analysis that continously estimates perceived motion in terms of `cumulative smooth pursuit'.  In essence, smooth pursuit segments are identified, interpolated where necessary, and joined probabilistically into a continuous record of `cumulative smooth pursuit'  (i.e., probability of eye position disregarding blinks, saccades, signal losses, and artefacts).  The analysis is fully automated and robust in healthy, developmental, and patient populations.   To validate reliability, we compare volitional reports of perceptual reversals in rivalry displays, and of physical reversals in non-rivalrous control displays.  `Cumulative smooth pursuit' detects physical reversals and estimates eye velocity more accurately than existing methods do \citep{frassle2014binocular}.  It also appears to distinguish dominant and transitional perceptual states, detecting changes with a precision of $\pm100\,\mathit{ms}$.   We conclude that `cumulative smooth pursuit' significantly improves the monitoring of binocular rivalry by means of recording OKN. 

\end{abstract}



\pagebreak

\section{Introduction}

Studies of binocular rivalry typically rely on volitional reports from observers  trained to communicate their subjective perceptual experience as rapidly and faithfully as possible \citep{logothetis1996rivalling,blake2002visual,tong2006neural,sterzer2009neural}.  Although subjective reports are perfectly adequate for numerous research questions, they suffer from certain limitations.  For example, volitional reports cannot be produced repeatedly at short intervals ($<200ms$), the underlying subjective criteria are difficult to establish, volitional reports require cooperative and healthy observers to be informative, and they necessitate additional neural activity, contaminating any activity associated with perceptual reversals.

An alternative to volitional reports are so-called `no-report' paradigms, which seek to monitor perceptual state on the basis of objective behavioural or physiological measures \citep{tsuchiya2015no,overgaard2016can}.   The most established `no-report' paradigm relies on optikinetic nystagmus (OKN), but modulation of pupil diameter and entrainment of physiological responses (EEG) by frequency- or contrast-tagged displays have also been widely used \citep{lansing1964electroencephalographic,brown1997method,tononi1998investigating,kornmeier2005necker,kamphuisen2008no,kornmeier2012eeg,jamison2015ssvep}.  

It has long been understood that OKN can reveal subjective perceptual experience, provided that the rivalrous displays are designed to elicit antagonistic nystagmus responses \citep{enoksson1963binocular,fox1975optokinetic}.  For example, if translational motion to the left and right is presented dichoptically to both eyes, the smooth pursuit phases of OKN will typically follow the perceptually dominant motion.  When perceptual dominance reverses, the direction of smooth pursuit typically reverses as well.  The validity of this approach was confirmed with magnetic scleral coils both in non-human primates trained to report their subjective experience and in human observers \citep{logothetis1990binocular,wei1998alternation} and was subsequently extended to  infrared eye trackers    \citep{watanabe1999optokinetic,naber2011perceptual,frassle2014binocular}.

Typically, the recorded eye velocity is processed and filtered to extract the slow (`pursuit') phase of OKN \citep{naber2011perceptual,frassle2014binocular,fujiwara2017optokinetic} and slow velocity is categorized in a binary fashion, such as to identify periods with a consistent direction of perceived motion (`dominance periods').  While this approach reliably identifies long dominance periods, shorter periods of either dominance or transition are more difficult to resolve.

Here we report an improved analysis that yields a continuous record of `cumulative smooth pursuit' (CSP).  This record consists of a sequence of eye velocity estimates (with confidence limits), which seamlessly joins pursuit periods and interpolated periods.  The sequence of velocity estimates can be parsed into distinct phases of `pursuit dominance' (with a typical mean duration $\sim2\,\mathit{s}$) and `pursuit transitions' (typical duration $\sim0.25\mathit{s}$ to $\sim0.5\mathit{s}$).  The change-over between phases can be determined with a precision of approximately $\pm100\,\mathit{ms}$.   

The analysis is robust and enables studies with large and diverse observer groups, including developmental cohorts, patient populations, and persons with idiosyncratic oculomotor patterns.   In control experiments with physically reversing image motion, ocular responses ({\it i.e.}, reversal of smooth pursuit direction) detected by CSP exhibited significantly less temporal variability (approximately 55\% smaller interquartile range) than ocular responses detected by existing methods \citep{naber2011perceptual,frassle2014binocular}.   During nearly linear pursuit episodes, CSP estimates proved approximately 15\% more accurate than existing methods.  In general, CSP analysis appeared to be marginally more sensitive and/or more volatile than existing methods, with estimates covering a slightly higher range of velocity, acceleration, and jerk.

We conclude that `cumulative smooth pursuit' offers significant improvements over existing methods of monitoring binocular rivalry by means of recording OKN.

\pagebreak

\section{Methods}
\subsection{Subjects}
The present study was performed on several observer cohorts, including neurotypical children (28, aged 12), adolescents (19, aged 16), young adults (30, average age 21), older adults (12, age 60), individuals with borderline disorder (12, average age 27), and individual with autism spectrum disorder (12, average age 28).  In addition, a number of more demanding control experiments were performed with practised psychophysical observers (2 young males, 6 young females, average age 24).

All participants had normal or corrected-to-normal vision, and were naive to the purposes of the experiment.  All observers passed a stereoacuity test before participating in the experiment (Super Stereoacuity Timed Tester, by Stereo Optical Co., U.S. Patent No. 5,235,361, 1993).  All participants or caregivers (in the case of children) provided informed written consent. For neurotypical observers, the study was approved by the Ethical Review Committee of the Institute of Psychology, Pazmany Peter Catholic University.  For the observes with autism spectrum disorder and borderline personality disorder, the study was approved by the Semmelweis University Regional and Institutional Committee of Science and Research Ethics.

The diagnostic status of participants with borderline personality disorder was assessed by the Hungarian version of the Structured Clinical Interview for DSM-IV Axis I and II disorders (\citep{unoka2004,unoka2006}).  Nine of twelve participants with autism spectrum disorder were diagnosed by a trained psychiatrist. They underwent a general psychiatric examination and their parents were interviewed about early autism specific developmental parameters. All nine participants fulfilled the diagnostic criteria of autism spectrum disorder, including autism specific signs between the critical ages of 4-5 years. Three of twelve participants in this group were recruited from a non-profit organization (Aura Organization) assisting people with autism spectrum disorder.  No detailed diagnostic information was available for these three participants.
\citep{unoka2004}
\subsection{Experimental stimuli and protocol}

\subsubsection{Setup}

\begin{figure}
	\centering
	\includegraphics[width=0.8\linewidth]{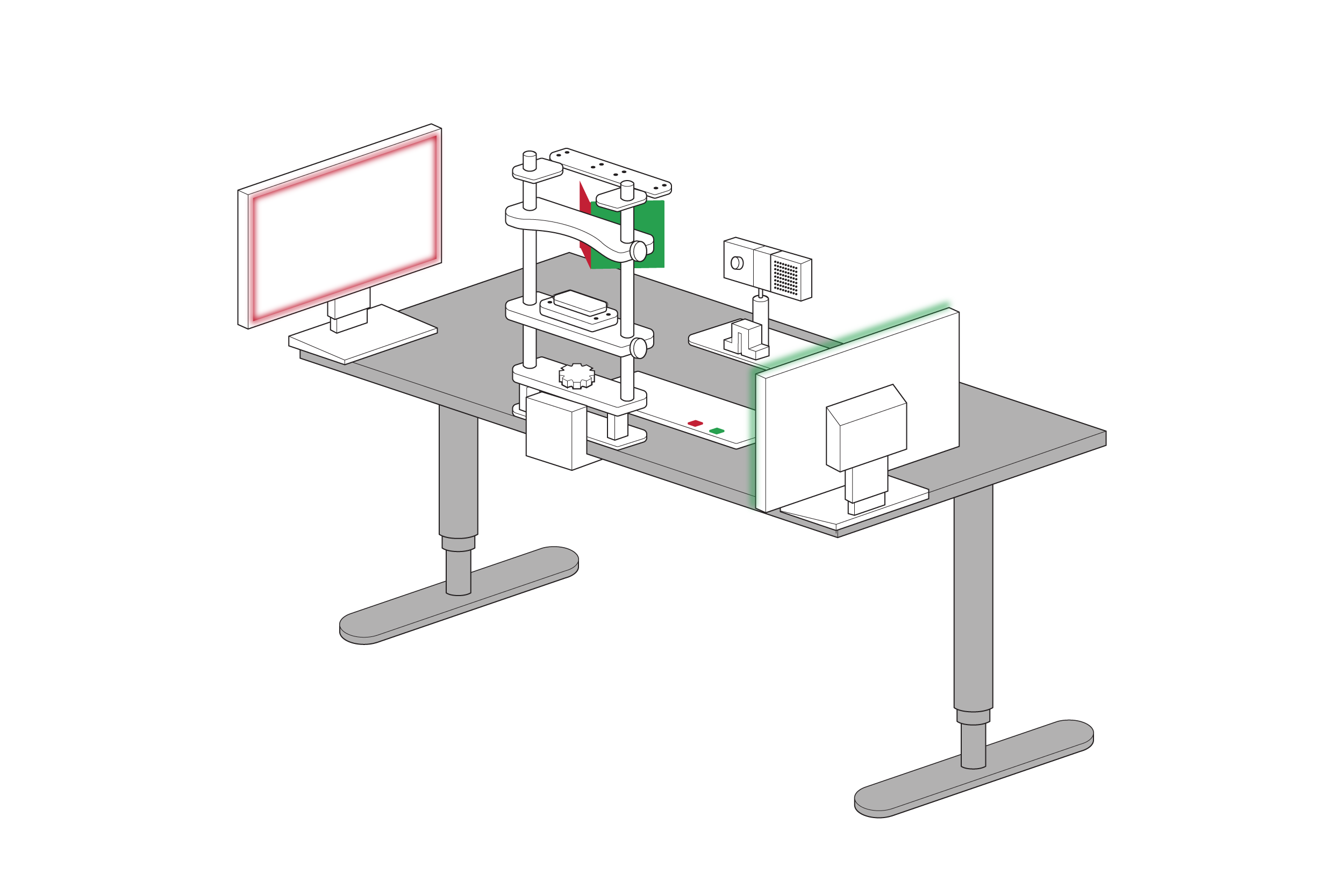}
	\caption[methods_fig]{\figtitle{Setup for dichoptic stimulation and eye position recording.} 
	Observers are immobilized by a headrest and view two displays through $45^\circ$ mirrors.  Each eye views a different display.  The camera for eye movement recording has a clear field of view, because the mirrors are transparent to infrared light.}
	\label{fig:setup}
\end{figure}

Nearly identical setups for dichoptic stimulation and eye position recording were used in Magdeburg and in Budapest, as described previously \citep{pastukhov2010rare}, (Fig.\,\ref{fig:setup}).  Observers were fixated by a headrest and view the two displays through $45^\circ$ mirrors, with each eye viewing a different display.  The mirrors are coated such as to reflect visible light but are transparent to infrared light, providing a clear field of view for the infrared camera that records eye position.  \cite{brascamp2017eye} recently described a rather similar setup in detail.

\subsubsection{Stimuli}

The stimuli consisted of green-and-black gratings for one eye, and red-and-black gratings for the other eye. The gratings moved horizontally, either leftward or rightward.  From the observer's point of view, the two gratings underwent uniform horizontal motion, either consistently (in the same direction) for the `replay condition', or inconsistently (in opposite directions) for the `rivalry condition'.  Each grating subtended a rectangular area of $15.2^\circ$ width and $8.4^\circ$ height.  Spatial frequency was $0.26\,\mathit{cycles}/^\circ$ and temporal frequency $8.7\,\mathit{cycles}/s$  horizontally. The speed of horizontal motion was $33.5\,^\circ/s$ or $1600\,\mathit{pix}/s$.  To facilitate binocular fusion, gratings were framed by a rectangular box with a random texture pattern. Stimuli was generated with Psychophysics toolbox 3 \citep{brainard1997psychophysics, pelli1997videotoolbox, kleiner2007s} running under Matlab R2015a. The spatial resolution was $48\,\mathit{pix}/^\circ$ the temporal refresh rate was $120\,\mathit{Hz}$.

\subsubsection{Protocol}

The experiment with developmental and patient cohorts consistent of ten trials of $95\,\mathit{s}$ duration.  The first trial served to familiarize observers with the display and was not included in the analysis.   It comprised  $20\,\mathit{s}$ of consistent grating motion in alternating directions,  $72\,\mathit{s}$ of inconsistent motion (which induced rivalry), finishing with $3\,\mathit{s}$ of consistent motion.  The remaining nine trials consistent of  $2\,\mathit{s}$ of consistent motion,  $92\,\mathit{s}$ of inconsistent motion, and $1\,\mathit{s}$ of consistent motion.  The consistent episodes served to reduce eye strain and to test the ocular response to physical motion reversals.  

The control experiments with experienced psychophysical observers consisted of six trials.  A first `familiarization' trial lasted $30\,\mathit{s}$ with consistent grating motion reversing every $3\,\mathit{s}$.   A second `passive viewing' trial lasted $90\,\mathit{s}$ with consistent grating motion reversing at random intervals sampled from a $\gamma$ distribution with mean $\mu = 3\,\mathit{s}$ and standard deviation $\sigma = 1.8\,\mathit{s}$.  

Under the `passive' conditions, observers were instructed to view the display as attentively as possible and to refrain from blinking as much as possible.

The four remaining `active viewing' trials lasted $90\,\mathit{s}$ and included two `replay' and two `rivalry' trials.  In `replay' trials, grating motion was always consistent and reversed at random intervals sampled from the $\gamma$ distribution describe above.  In `rivalry' trials, grating motion was consistent for  $2\,\mathit{s}$ and inconsistent for the remaining  $88\,\mathit{s}$ of the trial.  `Replay' and `rivalry' trials were presented in random order.  

Under `active' conditions, observers were instructed additionally to report the apparent direction of motion by pressing one of two keys.  Specifically, they were requested to report the initial apparent direction and any subsequent reversals of apparent direction by pressing the right- or left-arrow keys.

\subsection{Eye movement analysis}

Horizontal position of the left eye was recorded with an Eyelink 1000 (SR Research Ltd, Ottawa, Canada), with temporal sampling of $1\,\mathit{kHz}$.  The analysis of recordings involved several steps.  Briefly, fast and slow phases of horizontal OKN were identified and, after removing the former, the latter were interpolated into a continuous and cumulative record of smooth pursuit (CSP).  Below, the principal steps are described in more detail. 
\subsubsection*{Artefact removal and extraction of slow OKN phases}
Raw recordings are typically contaminated by blinking artefacts, in which nominal gaze position falls outside the display area.  Such off-scale events were removed, together with the adjoining $50\,\mathit{ms}$ on either side, from the recording.

To distinguish fast and slow segments of horizontal OKN, the recording was filtered bidirectionally with a $50\,\mathit{ms}$ kernel.  Specifically, a moving average was computed separately for $50\,\mathit{ms}$ windows sliding forward and backward in time.  The two moving averages were combined into a filtered record of eye positions.  Absolute eye velocity $|v|$ was computed numerically from successive position values. The velocity distribution was bimodal and the saddle point between the densities of low and high velocity was adopted as threshold criterion ($1.5\,\mathit{pix}/\mathit{ms}$).  Slow segments represented periods of smooth pursuit.  Fast segments represented fast OKN phases,  saccade-like eye movements, and occasional recording artefacts.  Positive or negative eye acceleration $a$ was computed from successive velocity values.  The distribution of accelerations peaked symmetrically at zero acceleration, with the $5\%$ and $95\%$ quantiles observed at approximately $\mp 0.12 \mathit{pix}/\mathit{ms}^2$, respectively.

\begin{figure}
	\centering
	\includegraphics[width=0.7\linewidth]{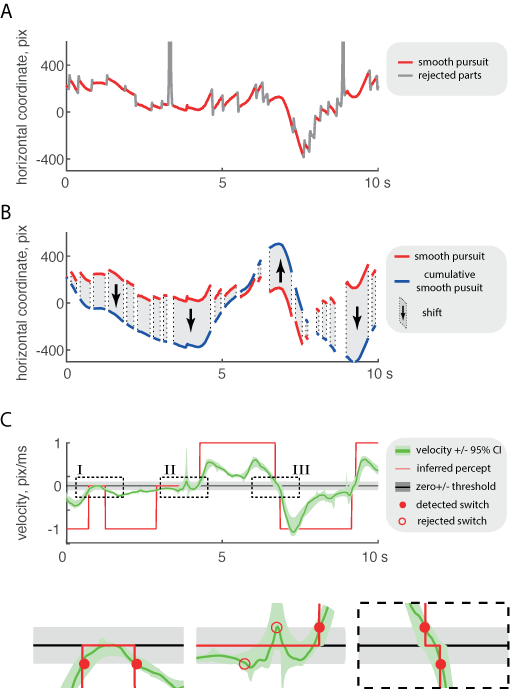}
	\caption[revision_method]{\figtitle{Processing of eye position record and detection of perceptual reversals.} 

		\textbf{(A)} Eye position record (gray trace) was parsed to identify pursuit segments (red traces). Smooth pursuit was defined by low velocity ($<$1.5 pix/ms) and low acceleration ($<\pm$95\% CI)
		\textbf{(B)} Extracted pursuit segments (red traces) were shifted vertically into alignment (blue traces), as indicated (arrows).  Shifted segments were interpolated and jointed by robust splining into a continuous probability density of CSP( `cumulative smooth pursuit', not shown).
		\textbf{(C)} Numerical differentiation of CSP density yields the density of CSP velocity, with mean and 95 \%CI as shown (green trace and light green area).   Perceptual state was inferred  from threshold-crossings (positive or negative) of the full confidence range (CI), as illustrated by insets.  
		\textbf{(I)} Return transition: CI crosses and recrosses one threshold (here, negative threshold).  Transition times (red dots) are the nearest threshold-crossing of mean velocity (green trace); \textbf{(II)} Rejected transitions of negative and positive threshold (red circles) and accepted transition of positive threshold (red dot). \textbf{(III)} Forward transition: CI crosses both positive and negative threshold.  Transition times (red dots) are the nearest threshold-crossing of mean velocity (green trace).}
	\label{fig:methodfig}
\end{figure}

Time segments of slow velocity ($|v| \leq 1.5\,\mathit{pix}/\mathit{ms}$) {\it and} low acceleration \\($|a| \leq 0.12\, \mathit{pix}/\mathit{ms}^2 $) were retained, provided their duration exceeded $50\,\mathit{ms}$ (Fig.\,\ref*{fig:methodfig}A, red traces).  All other segments were disregarded (Fig.\,\ref*{fig:methodfig}A, gray traces). 

Importantly, all subsequent analyses (including concatenation and interpolation) were based on retained time segments from the original, raw recording (rather than the bidirectionally filtered recording).  Thus, the final result did {\it not} rely on filtered eye position records.

\subsubsection*{Continuous record of smooth pursuit}
In the present context, eye velocity is in the centre of interest, not absolute eye position.  Accordingly, we shifted the absolute position of each segment vertically, such as to maintain continuity of both velocity and position, after shifting.  Due to these positional shifts, absolute position was replaced by cumulative position.  Specifically, given successive smooth pursuit segments $x_1(t)$, $t\in T_1$, and $x_2(t)$, $t\in T_2$, we joined these segments by fitting a four-parameter function $f(t)$ (three-parameter parabola plus offset) to the final $50\,\mathit{ms}$ of the earlier and the first $50\,\mathit{ms}$ of the later segment:
\ba
f(t)= 
\begin{cases}
at^2+bt+c,& \text{if } t\in T_1\\
at^2+bt+c-\mathit{offset},& \text{if } t\in T_2\\
\end{cases}
\ea
Figure  \ref{fig:methodfig}B  shows a representative recording with unshifted segments (red curves) and shifted segments (blue curves) corresponding to the best parabolic fit.  After shifting, the segments form a cumulative, but still intermittent, record of smooth pursuit (CSP).  Note that positive CSP slope corresponds to rightward pursuit, negative slope to leftward pursuit. 

To obtain a continuous estimate of positive or negative eye velocity $v$, the disjoint segments of CSP were used as anchors for `robust splining' ({\it i.e.}, repeated splining of random subsamples).  Specifically, each recording was subsampled $10^3$ times by a factor of $1 /100$ and splined with a shape-preserving, piecewise-cubic, Hermite-interpolating polynomial (PCHIP, \cite{carlson1985monotone}).  The splined subsamples were averaged and the time-derivative was computed numerically.  The median and the 95\%CI of estimated eye velocity are illustrated in Figure \ref{fig:methodfig}C (green trace and light green area).   The same three quantiles ($2.5\%$, $50\%$ and $97.5\%$) of the velocity distribution provided the basis for estimating perceived motion (see next section). 

To summarize, the present approach (i) smoothes raw observations in an adaptive manner (by robust splining), (ii) interpolates gaps in a manner that avoids discontinuities in velocity (by splining), and (iii) estimates local velocity and acceleration from a neighborhood (a splined segment).

\subsubsection*{Estimate of perceived motion}

As horizontal smooth pursuit tends to follow perceived horizontal motion \citep{enoksson1963binocular,merrill1963optokinetic,fujiwara2017optokinetic,frassle2014binocular}, we inferred perceived motion from horizontal velocity of smooth eye movements.  In principle, zero-crossings of horizontal eye velocity may indicate reversals of perceived motion.   However, horizontal eye velocity may approach zero for several reasons other than a reversal of perceived motion, including slowing of pursuit, lapses of perception or attention, transitions from and to the same perceived motion (``return transition''), perception of mixed motion, and artefacts introduced by processing.

To distinguish conclusive from marginal zero-crossings, we used a gaze velocity threshold of $\pm 0.1 \mathit{pix}/\mathit{ms}$ as an additional criterion (Fig.\,\ref{fig:methodfig}{C}, gray area; Fig.\,\ref{fig:result_figure}{A}, gray vertical stripe).  This value was a conservative choice and larger values, up to approximately $\pm 0.4 \mathit{pix}/\mathit{ms}$, would have served equally well.

Conclusive transitions were defined in terms of the entire velocity confidence interval (from $2.5\%$ to $97.5\%$ quantiles) crossing either the upper threshold (from above or below) or the lower threshold (from above or below). The timing of such a conclusive transition was defined by the nearest threshold-crossing of the $50\%$  quantile.  This approach defined both the beginning and the end of `dominance phases' and `transition phases'.  It also allowed us to distinguish different kinds of `transition phases', specifically, `forward transitions' leading to the opposite dominance as previously and `return transitions' leading to the same dominance as previously (see Fig.\,\ref{fig:methodfig}, insets I and III, respectively).   Note that this approach rejects marginal transitions without resorting to a temporal criterion (see Fig.\,\ref{fig:methodfig}, inset II).  Accordingly, the lower bound for `dominance' and `transition' durations is not set explicitly but implicitly by the confidence interval for velocity. 

The precision of the determination of the beginning or ending of a transition phase was estimated individually for each threshold-crossing of the average spline.  To this end, we computed the standard deviation of threshold-crossings (of the time-derivative of) individual splines around the threshold-crossing of the average spline.  The resulting value was essentially the half-width of a $67\%$ confidence interval.

\subsection{Method of Fr\"assle and colleagues (2014)}

To compare our results to existing methods, we implemented the algorithm of Fr\"assle and colleagues (2014).  To this end, eye velocity was computed numerically from successive values of horizontal eye position.  High-frequency noise was suppressed by averaging over a $500\mathit{ms}$ sliding temporal window.   Gaps in the eye velocity record were interpolated linearly.  To assess perceptual state, zero-crossings in the filtered and interpolated eye velocity were determined.  Specifically, zero-crossings were retained only if separated by $400\mathit{ms}$ or more.  More closely spaced zero-crossings were disregarded.  Thus,  $400\mathit{ms}$ was the minimal duration of detected perceptual states.  

\subsection{Statistical methods}

\subsubsection*{Summary statistics}

Unless otherwise mentioned, we report grand means of data pooled from all observers, plus minus the standard error of the mean (SEM).   Distributions were compared in terms of differences between medians and/or between interquartile ranges (IQR) or between $95\%$ confidence intervals (CI, interquantile range from $2.5\%$ to $97.5\%$ ).

\subsubsection*{Combined distributions}

To highlight similarity of distributions obtained from different observers, it was sometimes useful to first normalize observations for each individual observer.

For example, we normalized CSP velocity distributions of individual observers by means of z-scoring (normalizing to zero mean and unit variance), prior to pooling all observations in a combined distribution of velocity (Fig.\,\ref{fig:result_figure}A).   Note that the low-velocity range $\pm 0.1 \mathit{pix}/\mathit{ms}$ occupied a slightly different position for each observer in terms of z-score units.

Similarly, we normalized dominance durations to the individual observer mean $\expect{T_\mathit{dom}}$, prior to pooling all observations in a combined distribution of dominance durations (Fig.\,\ref{fig:result_figure}D).

\subsubsection*{Statistical tests}
To assess the statistical significance of differences between the medians of distributions of latencies (Fig.\,4 BC) or durations (Fig.\,8D), we used the Wilcoxon rank-sum test.  

To assess the statistical significance of differences between the proportions of return transitions (Fig.\,8C), we used the z-test for binomial distributions. 

To assess the statistical significance of differences between interquantile ranges (Fig.\,4BC), we resorted to a bootstrapping approach.  The two original sets of samples were merged and randomly divided into two new sets.  This sampling was repeated ($>10^4$ times) and a distribution of interquantile ranges was established (null hypothesis).  The significance of observed interquantile ranges was then assessed in terms of this null distribution.

\subsubsection*{Violin plots}
To visualize the distributions of latencies and durations (Figs.\,4 and 8), we used `violin plots' implemented by the gramm MatLab toolbox \citep{morel2018gramm}. These plots superimpose a normal `box plot' and a violin-shaped rendering of a distribution. 

Box plots present the median $\pm$ interquartile range (IQR) of the distribution.  Notches were drawn at $\pm 1.58*IQR/\sqrt{N}$,where N is number of observations. The whiskers extend above and below the box to the most extreme data points that are within a distance to the box equal to 1.5 times the interquartile range (Tukey boxplot). Points outside the whisker range were plotted as outliers.  The violin shape visualizes the probability density function and represents the distribution of samples.  The thickness of the violin indicates how common (probable) a given sample value is.

\pagebreak
\section{Results}

\begin{figure}
	\centering
	\includegraphics[width=0.8\linewidth]{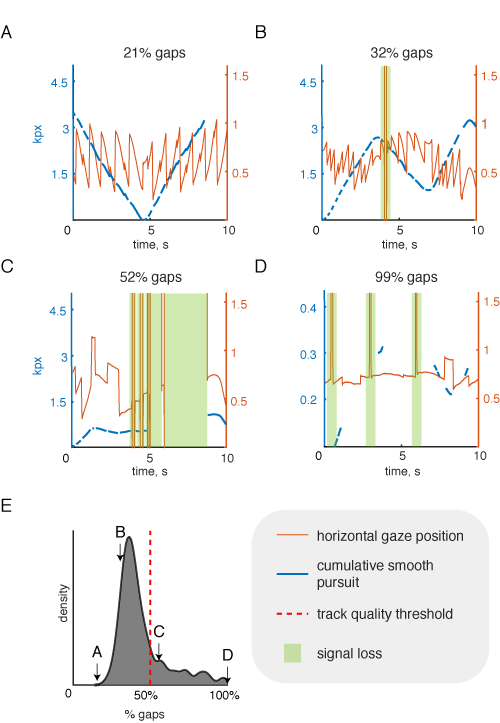}
	\caption[quality_examples]{\figtitle{Examples of eye position records.}  Raw records of horizontal eye position (red traces), periods of signal loss (green areas), and extracted segments of `cumulative smooth pursuit' (blue traces), both in units of kilo-pixel ($\mathit{kpx}$).  The respective proportions of pursuit segments and gaps determines the quality of a record.  \textbf{(A-B)} illustrates `good' records included in further analysis.  \textbf{(C-D)} illustrates `poor' records excluded from further analysis.   \textbf{(E)}  Distribution of  quality for $983$ records.  The exclusion threshold was 50\% (red dashed line). }
	\label{fig:qualityexamples}
\end{figure}

To assess the performance of the `cumulative smooth pursuit' (CSP) analysis, we recorded eye position during binocular rivalry in 30 neurotypical observers of age 21 years and compared the results to existing analysis methods \citep{frassle2014binocular}.   To assess the robustness of CSP for different populations, we also investigated 59 neurotypical observers of age 12, 16, and  60 years, as well as a further 24 observers with borderline or autism spectrum disorder.  Note that statistical differences in binocular rivalry as experienced by different age and patient groups are not relevant in the present context and will be reported elsewhere.

To assess the reliability of `cumulative smooth pursuit' as an indicator of perceived motion, we compared volitional reports of perceived motion in two control experiments with an additional group of 8 trained psychophysical observers.  One experiment investigated ocular responses to physical reversals of image motion in non-rivalrous displays (`replay condition') and the other ocular responses associated with spontaneous perceptual reversals induced by binocular rivalry displays (`rivalry condition').  Additionally, both experiments were performed under both `active viewing' conditions ({\it i.e.}, with volitional reports) and `passive viewing' conditions ({\it i.e.}, without such reports).

\subsection{Robustness of CSP analysis}
To assess the robustness of CSP analysis, we processed $983$ eye position recordings of $90\,\mathit{s}$ duration, from $113$ neurotypical observes of different ages (12, 16, 21, and 60 years) and from $24$ individuals with borderline disorder (BD) or autism spectrum disorder (ASD).

Eye movement records were analyzed in a fully automated fashion, as described in Methods.  Four examples of rivalry recordings of different quality, all from healthy, 21 year-old observers, are illustrated in {\bf Figure~\ref{fig:qualityexamples}}.  The original eye position record (red trace) including eye blinks (green overlay)  is compared to CSP segments (blue traces).  Positive CSP slope corresponds to rightward pursuit, negative slope to leftward pursuit. The quality of each record was assessed in terms of the relative proportion of identified smooth pursuit segments and of gaps between such segments.  The fraction of the former ranged from 79\% to 1\%. Recordings with at least 50\% identified smooth pursuit segments were included in further analyses, all other recordings were disregarded {\bf Figure~\ref{fig:qualityexamples}E}.

In our corpus of $983$ recordings, $804$ recordings (82\%) exceeded the quality threshold.  Of $113$ observers, $99$ observers (90\%) produced recordings of acceptable quality at least half the time.   The remaining 10\% of observers exhibited unusually slow velocities of smooth pursuit, which were reduced more than ten-fold as compared to the other observers ({\bf Fig.~\ref{fig:qualityexamples}D}).  It is unclear whether this unusually low gain reflected  measurement problems or genuine physiological deviations. 

\begin{figure}[hb!]
	\centering
	\includegraphics[width=1\linewidth]{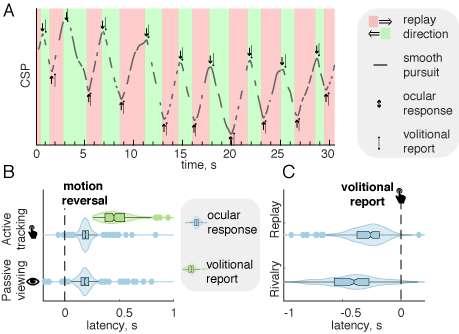}
	\caption[example_track]{\figtitle{Relative latency of ocular response and volitional report.}  {\bf (A)}  Example of processed recording of `replay condition'.  Shifted segments of smooth pursuit, CSP (gray traces), compared to physical reversals of horizontal image motion (green and red stripes).  Arrows indicate the time of ocular responses (thick) and volitional reports (thin).  {\bf (B)} Latency of ocular responses (blue violin plots) and volitional responses (green violin plots), relative to physical reversal of horizontal image motion, during active and passive viewing of `replay' display.   {\bf (C)} Latency of ocular responses (blue violin plots), relative to volitional report, during active viewing of `replay' and `rivalry' displays.}
	\label{fig:exampletrack}
\end{figure}

\subsection{Relative latency of CSP analysis}

To assess the reliability of CSP as an indicator of perceived motion, we compared reversals of smooth pursuit direction, to reversals of physical image motion and to volitional reports of reversals of physical and/or perceived motion, using both healthy young adults and practised psychophysical observers.  

Under `rivalry' conditions, observers dichoptically viewed a rivalrous display, in which perceived motion reversed spontaneously at irregular times.  Under `replay' conditions, observers binocularly viewed a non-rivalrous  ({\it i.e.} unambiguous) display, in which image motion reversed physically at irregular times.   Frequency and variability of reversal timing was comparable under both conditions (hence `replay').  Eye position records were analyzed as described above in terms of CSP.  Reversals in the direction of smooth pursuit were interpreted as `ocular responses' to reversals of physical or perceived image motion.   

In addition, observers produced `volitional reports' of perceived image motion.  Specifically, observers were instructed to press either a ``motion left'' or a ``motion right'' key, whenever the perceived motion changed.  In the `replay' condition, `ocular responses' and `volitional reports' could also be compared to reversals of physical image motion.  

In addition to an `active viewing' condition in which observers produced `volitional reports', we also investigated a `passive viewing' condition in which observers simply viewed the display.   Unsurprisingly, observers appeared to be somewhat less alert and attentive under `passive viewing' conditions.

Overall, we recorded eye position under `replay \& passive viewing' conditions ($16\,\mathit{min}$, $512$ reversals total), `replay \& active viewing' conditions ($24\,\mathit{min}$, $727$ reversals total),  and `rivalry \& active viewing' conditions ($24\,\mathit{min}$, $718$ reversals total).
For `rivalry \& passive viewing' conditions, we used part of the larger data set mentioned above (30 neurotypical observers of age 21, $25\,\mathit{h}$ and $40418$ reversals total).

A representative example of a recording with reversals of physical image motion (`motion reversal'), reversals of smooth pursuit direction (`ocular response'), and reported reversal of perceived motion (`volitional report') is shown in Figure \ref{fig:exampletrack}A. 


The latencies between the three types of events -- motion reversal, ocular response, volitional report --  under various conditions are summarized in Figure~\ref{fig:exampletrack}BC and in Table~1.  When  `replay' displays were viewed actively, reversals of image motion were followed almost invariably by a corresponding reversal of ocular motion ($95\%$, latency $< 1\mathit{s}$) and a volitional report ($95\%$, latency $< 1\mathit{s}$).   Relative to motion reversals, the mean latency of ocular responses was $179\mathit{ms}\pm{7.5}\,\mathit{(SEM)}$ and that of volitional responses was $454\mathit{ms}\pm{9.8}$ (Fig.\,\ref{fig:exampletrack}B).  The difference between medians was highly significant ($p< 10^{-3}$). The mean interval between ocular responses and volitional reports was $259\mathit{ms}\pm48$.  When the same display was viewed passively, the latency of ocular responses was $189\mathit{ms}\pm{9}$, which did not differ significantly from active viewing.

The respective variability of ocular responses and volitional reports was quite different.   Specifically, when `replay' displays were viewed actively, we obtained $95\%$ interquantile ranges of $97.5\,\mathit{ms}$ for ocular responses and of $151\,\mathit{ms}$ for volitional reports.  This difference was significant ($p<0.05$).  The variability of intervals between ocular responses and volitional reports was $326\,\mathit{ms}$ ($95\%$ interquantile range).


When `rivalry' displays were viewed actively, the mean interval between ocular responses and volitional reports was $400\mathit{ms\pm53} \mathit{SEM}$ (Fig.\,\ref{fig:exampletrack}C).  Individual intervals were highly variable, with a $95\%$ interquantile range of $393\,\mathit{ms}$.  Both mean and  variability of these intervals were significantly larger than for `replay displays', as described above ($p<.01$ and $p<.001$, respectively).  There could be several possible reasons for this difference, including that reversals of perceived motion proceed more gradually and/or more variably than reversals of physical motion.


{\renewcommand{\arraystretch}{2}%
	\begin{table}[ht]
		\caption{Relative latency of motion reversal (MR), ocular response (OR), and volitional report (VR).  The coloring corresponds to {\bf Fig.\,\ref{fig:exampletrack}BC}.} 
		\centering 
		\begin{tabular}{c|cc|c|c}
			\cline{2-4}
             & \multicolumn{2}{c|} {Replay}   & Rivalry        &       \\ 
             \cline{2-4} 
	         & From MR \ldots                       & From OR \ldots     & From OR \ldots    &    \\ 
	         \hline
			\multicolumn{1}{|c|}{} & \cellcolor[rgb]{0.62,0.76,0.8}$179\pm75\,\mathit{ms}$  & n.a. & n.a. & \multicolumn{1}{c|}{\ldots to OR} \\
			\cline{2-2}
			\multicolumn{1}{|c|}{\multirow{-2}{*}{Active}} & \cellcolor[rgb]{0.66,0.82,0.55}$454\pm98\,\mathit{ms}$ & 
			\cellcolor[rgb]{0.62,0.76,0.8}$275\pm49\,\mathit{ms}$  & \cellcolor[rgb]{0.62,0.76,0.8}$400\pm53\,\mathit{ms}$ 
			& \multicolumn{1}{c|}{\ldots to VR} \\ 
			\hline
			\multicolumn{1}{|c|}{Passive} & \cellcolor[rgb]{0.62,0.76,0.8}$189\pm9\,\mathit{ms}$ &n.a. &n.a. & \multicolumn{1}{c|}{\ldots to OR}\\
			\hline
		\end{tabular}
	\end{table}

\subsection{Comparison to existing methods}

To compare CSP results to existing methods \citep{naber2011perceptual,frassle2014binocular}, we re-analyzed our eye position recordings with the algorithm of Fr\"assle and colleagues (2014).   For active viewing of `replay' displays, the latency of ocular responses (relative to motion reversal) detected by the two approaches was statistically indistinguishable, with a median of median $189 \,\mathit{ms}$ for CSP and $183 \,\mathit{ms}$ for Fr\"assle and colleagues.   However, the variability of latencies was considerably smaller for CSP (interquartile range $55\,\mathit{ms}$) than for the method of Fr\"assle and colleagues (interquartile range $123\,\mathit{ms}$).  The difference in variability was highly significant ($p<10^{-6}$).  Accordingly, CSP results were approximately $55\%$ less variable than existing methods.

In a second comparison, we assessed to precision of eye movement estimates from CSP and from the algorithm of Fr\"assle and colleagues (2014).  For this purpose, we focussed on highly linear smooth pursuit episodes in our recordings from 30 neurotypical observers (age 21).  For each such episode, we estimated true eye velocity as the slope of a fitted regression line (with $r>0.999$) and analyzed the distribution of residuals (difference between CSP velocity estimates and `true' velocity).  For episode durations from $150\,\mathit{ms}$ to $550\,\mathit{ms}$, average residuals ranged from $45\,\mathit{pix}/\mathit{ms}$ to $80\,\mathit{pix}/\mathit{ms}$ (mean $56 \, \mathit{pix}/\mathit{ms}$).   With the method of Fr\"assle and colleagues, average residuals ranged from  $57\,\mathit{pix}/\mathit{ms}$ to $82\,\mathit{pix}/\mathit{ms}$ (mean $66 \, \mathit{pix}/\mathit{ms}$).  Although this difference was not large (approximately $15\%$), it was highly significant ($p<10^{-6}$).  Accordingly, even under ideal conditions of highly linear eye motion, CSP analysis further improves the precision of (already good) eye velocity estimates.

In a third comparison, we considered the distribution of absolute eye velocity $|v|$, absolute acceleration $|a|$, and absolute jerk $|j|$ (change in acceleration), as estimated by CSP and by the algorithm of Fr\"assle and colleagues (2014) from 30 neurotypical observers (age 21).  As described in Methods, the CSP analysis smoothed and interpolated the eye position record in an adaptive (context-dependent) manner.  The median and interquartile range of the resulting distributions were $440\pm230\mathit{pix}/\mathit{s}$ for $|v|$,  $4850\pm4125\mathit{pix}/\mathit{s}^2$ for $|a|$, and  $7.4\pm6.7\times10^{5}\mathit{pix}/\mathit{s}^3$ for $|j|$.  In contrast, Fr\"assle and colleagues (2014) computed eye velocity numerically from successive position values and filtered the result in a non-adaptive (context-insensitive) manner.  With this approach, the resulting median and interquartile range were $420\pm210$ for $|v|$,  $4500\pm3600$ for $|a|$, and  $5.0\pm3.6\times10^{5}$ for $|j|$, in the same units as above.   Thus, the median of velocity, acceleration, and jerk from the CSP analysis was consistently higher ($5\%$, $8\%$, and $25\%$, respectively), although the difference was significant only with respect to jerk ($p<0.2$, $p<0.1$, $p<0.05$, respectively).  

We suspect that this difference reflects a (desirable) increase in the sensitivity of eye velocity estimates.  However, we cannot rule out an (undesirable) increase in the volatility of estimates.  Deciding the issue would require a situation in which true eye velocity was known.


\subsection{Trajectory of CSP}

Before and after smooth pursuit reverses direction, the trajectory of CSP is rather stereotypical and consistent.  These average trajectories are of interest, as they track the progress of reversals of perceived motion under different conditions.

Average CSP trajectories during pursuit reversal are illustrated in (Fig.\,\ref{fig:cumulativesmoothpursuit}).  To obtain this figure, we temporally aligned several hundred individual trajectories, recorded from eight practised psychophysical observers, to obtain the mean eye position and its standard deviation, at different times before and after a reversal.  During active viewing of a `replay' display, average CSP trajectories develop almost monotonically before and and after reversals of image motion ('motion reversal'), pursuit direction (`ocular response'), and 'volitional report' (red traces, Fig.\, \ref{fig:cumulativesmoothpursuit}A).  During active viewing of a `rivalry' display, average CSP trajectories are less consistent, presumably reflecting the more gradual nature of perceptual reversals  (gray traces, Fig.\,\ref{fig:cumulativesmoothpursuit}A).   During passive viewing of `replay' displays, average trajectories again develop monotonically  (red traces, Fig.\,\ref{fig:cumulativesmoothpursuit}B).  However, during passive viewing of `rivalry' displays, cumulative trajectories again become less consistent   (gray traces, Fig.\,\ref{fig:cumulativesmoothpursuit}B).  It is possible that the greater variability of perceptual reversals is exacerbated by reduced alertness or attention under conditions of passive viewing.

\begin{figure}[h!]
	\centering
	\includegraphics[width=1\linewidth]{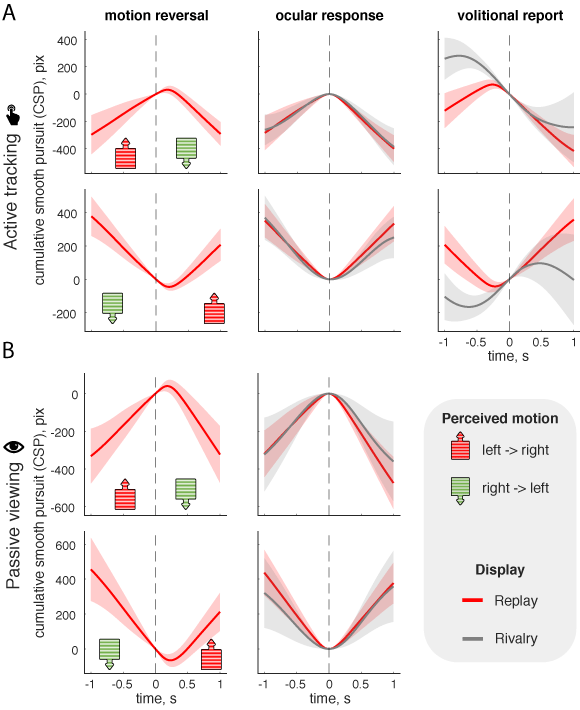}
	\caption[cum_pursuit]{\figtitle{Cumulative trajectories of smooth pursuit}.   Cumulative trajectories of smooth pursuit were temporally aligned to `motion reversals' (left column), `ocular responses' (middle column), and `volitional reports' (right column).  Rising trajectories represent rightward motion (red icon), falling trajectories leftward motion (green icon).  Average and standard deviation are shown separately for leftward-to-rightward and rightward-to-leftward reversals. {\bf (A)} Active viewing of replay (red) or rivalry (gray) displays.  {\bf (B)}  Passive viewing of replay (red) or rivalry (gray) displays.}
	\label{fig:cumulativesmoothpursuit}
\end{figure}

\clearpage

To take a closer look at the apparent differences between `replay' and `rivalry' conditions ({\it i.e.} between reversals of physical and perceived motion), we compared the respective acceleration of cumulative smooth pursuit, before and after it reverses direction (Fig.\,\ref{fig:accelerationfig}A).   During physical reversals, acceleration peaks sharply at the precise moment of the reversal, falling to near zero within a quarter second before and after (red traces).  In contrast, during perceptual reversals, acceleration rises and falls more gradually (gray traces).   The difference in peak acceleration is highly significant ($p<.0001$, Fig.\,\ref{fig:accelerationfig}B).  We suspect that this difference may reflect the more gradual and variable nature of perceptual reversals.   The extent to which cumulative pursuit can reveal the internal dynamics of perceptual reversals is discussed below.

\begin{figure}[h!]
	\centering
	\includegraphics[width=1\linewidth]{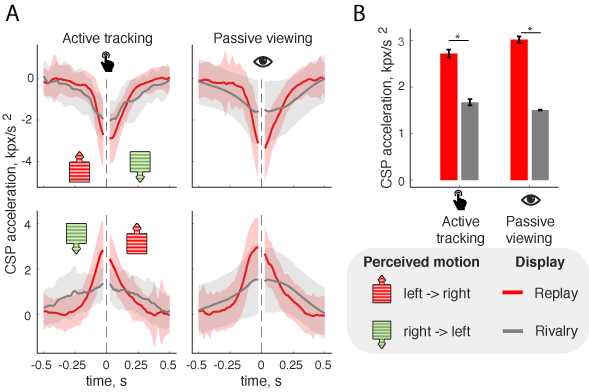}
	\caption[acceleration]{\figtitle{Acceleration of cumulative smooth pursuit, before and after reversal of direction.} {\bf (A)} Second derivative of cumulative smooth pursuit (in units of kilo-pixel, $\mathit{kpx}/\mathit{s}^2$) for active and passive viewing of `replay' and `rivalry' displays.  Zero marks the reversal of pursuit direction.  Average and standard deviation are shown separately for leftward-to-rightward and rightward-to-leftward reversals.  The central $50 \,\mathit{ms}$ reflect the interpolation algorithm (rather than ocular motion) and were excluded.  {\bf (B)}   Mean peak acceleration ($\pm$SEM), computed as the difference of velocity $-26\,\mathit{ms}$ before and $+26\,\mathit{ms}$ after the reversal pursuit direction. `*' indicates a p-value$<0.0001$,}
	\label{fig:accelerationfig}
\end{figure}

To assess the consistency of CSP trajectories in different observer groups, we repeated the analysis separately for different age and patient groups.  Qualitatively, almost all groups produced consistent results, with comparable pursuit velocities and accelerations before, during, and after  perceptual reversal (Fig.~\ref{fig:ambigswitches}).  The one exception was older observers (age 60), in whom pursuit velocity and acceleration were noticeably diminished. 

\begin{figure}
	\centering
	\includegraphics[width=1\linewidth]{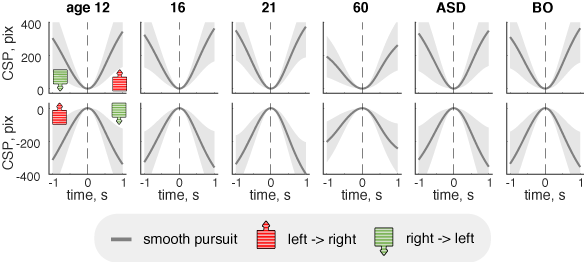}
	\caption[ambig_switches]{\figtitle{Average CSP trajectories for different observer groups}. Average CSP trajectories, before and after reversal of direction, for different observer groups, including healthy subjects of \textbf{12}, \textbf{16}, \textbf{21}, and \textbf{60 years of age}, and individuals with autism spectrum disorder, \textbf{ASD}, and borderline disorder, \textbf{BO}.   The horizontal gaze position at the detected time of reversal was set to zero.  Average and standard deviation are shown separately for leftward-to-rightward and rightward-to-leftward reversals.}
	\label{fig:ambigswitches}
\end{figure}

\subsection{Phases of smooth pursuit}

The approach described here reveals considerable detail about different phases of smooth pursuit behaviour.  The distribution of CSP velocity consistently exhibits two peaks for positive (rightward) and negative (leftward) pursuit, plus an intermediate peak at velocities near zero.  The combined distribution from 30 healthy observers (age 21 years) is shown in Figure \ref{fig:result_figure}A, in z-score units (see Statistical methods).  Clearly, pursuit velocity does not simply alternate between extended periods of positive or negative values, but also lingers for extended periods in a near-zero velocity regime.

Phases of strongly positive or negative pursuit velocity we term `pursuit dominance', as they presumably correspond to phases of {\it perceptual} dominance of rightward or leftward motion, respectively.   Phases of near-zero pursuit velocity we term `pursuit transition', as they intervene between `dominant' phases.   Which kinds of perceptual states correspond to such `transition' phases is at this point unclear and deserving of further study.  Presumably, pursuit `transitions' overlap to some degree with the perceptual states sometimes described as `mixed' or `patchy' rivalry \citep{brascamp2006time, pastukhov2011cumulative}. 

The parsing of CSP velocity into distinct phases of pursuit dominance and of pursuit transitions is illustrated  (Fig.\,\ref{fig:methodfig}C).  Importantly, `forward' transitions (leading to the opposite dominance) and `return' transitions (leading to the same dominance) may also be distinguished.
The approximate beginning and ending of such phases may be timed by comparing the confidence range of interpolated CSP velocity with a suitable low-velocity threshold (here $\pm 0.1\,\mathit{pix}/\mathit{ms}$). The precision of this timing may also be estimated from the local CSP confidence range (see Methods).  Overall, the average precision was approximately $50\,\mathit{ms}$ (see also below).

The combined results from 30 healthy observers (age 21 years) are summarized in Figure \ref{fig:result_figure}B--E and Figure \ref{fig:result_transition}. The observed `pursuit dominance' and `pursuit transition' durations were distributed approximately log-normally, as indicated by the approximately symmetric shape of their distribution on a logarithmic scale (Fig.\,\ref{fig:result_figure}D).  Dominance durations were far longer than transition durations.  Note that, for these combined distributions, individual observer distributions were normalized to the individual average value of $\left\langle T_\mathit{dom}\right\rangle$ (see Statistical methods).  Absolute values are given below.


Average duration of pursuit dominance was $\expect{T_\mathit{dom}}=1.90\mathit{s}\pm0.02\,\mathit{SEM}$.  The average duration of forward pursuit transitions was $\expect{T_\mathit{forward}}=230\mathit{ms}\pm 20$, whereas the average duration of return pursuit transitions was  $\expect{T_\mathit{return}}=530\mathit{ms}\pm 70$ (Fig.\,\ref{fig:result_figure} C).  The beginning and ending of forward (pursuit) transitions was determined with a precision of $50\mathit{ms} \pm 1$ (halfwidth of confidence interval).  For return transitions, this value was $80\mathit{ms} \pm 3$ (Fig.\,\ref{fig:result_figure} C).

Trajectories of eye velocity during pursuit transitions are summarized in Figure \ref{fig:result_transition}.
During long transitions, an extended period of constant low velocity is particularly evident (Fig.\, \ref{fig:result_transition}AC).  The extent to which velocity remains constant over all recorded transitions is visualized by plotting acceleration ({\it i.e.}, velocity difference before and after the midpoint) against transition duration (Fig.\, \ref{fig:result_transition}BD).  It is evident that, for transition durations over $200\mathit{ms}$, velocity typically remains nearly constant around the midpoint of the transition.

Are pursuit transitions a spurious phenomenon, due perhaps to temporary oculomotor indecision?  Or do pursuit transitions reflect the dynamics of {\it perceptual} transition states, such as `mixed' or `patchy' percepts \citep{brascamp2006time, pastukhov2011cumulative}?   To address these questions, we investigated the effect of perceptual adaptation on pursuit transitions, as this factor is known to interact with perceptual transitions \citep{pastukhov2011cumulative}.  To identify periods of comparatively weak or strong adaptation, we selected exceptionally `short' or `long' dominance periods $T_\mathit{dom}$ (where `short' and `long' were defined in therms of a $5\%$ and a $95\%$ quantile, respectively) (Fig.\,\ref{fig:result_figure}B). 

Return (pursuit) transitions were significantly more frequent immediately following short $T_\mathit{dom}$ ($11.5\%\pm1.0$) than immediately following long $T_\mathit{dom}$ ($6.0\%\pm0.9$).  The overall probability of return (pursuit) transitions was $7.9\%\pm0.9$.   Both return and forward (pursuit) transitions were significantly {\it longer} immediately after short than after long $T_\mathit{dom}$ (Fig.\,\ref{fig:result_figure} E).  This effect was highly significant for forward (pursuit) transitions (which were more frequent).   In contrast, dominance phases were significantly {\it shorter} after short than after long $T_\mathit{dom}$, consistent with a positive sequential correlation.

These results mirror the known effects of perceptual adaptation on {\it perceptual} transition and dominance phases (see Discussion), suggesting that {\it pursuit} transitions may reflect the underlying dynamics of {\it perceptual} transitions.

\begin{figure}
	\centering
	\includegraphics[width=0.7\linewidth]{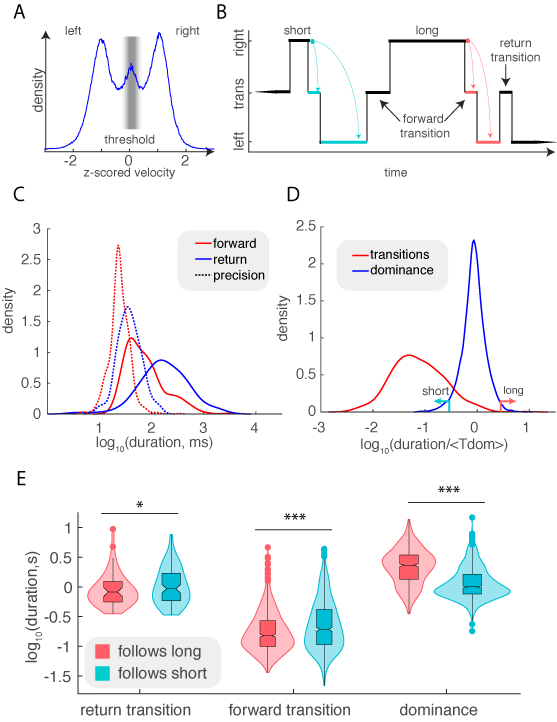}
	\caption[revision_results]{\figtitle{Phases of pursuit dominance and transition.} 
       \textbf{(A)} Trimodal distribution of CSP velocity for 30 observers, in z-score units (blue curve), and low velocity range of $\pm 0.1 \mathit{pix}/\mathit{ms}$ (blurred vertical bar).  Blurring indicates that, in z-score units, the low velocity range differs slightly between observers.
		\textbf{(B)} Schematic sequence of perceptual reversals, with upper and lower steps representing pursuit dominance and middle steps representing pursuit transition.  Phases following an exceptionally long (short) dominance period are marked in red (blue).
		\textbf{(C)} Probability density of forward and return transition durations (solid red and blue) and of the precision (halfwidth of confidence interval) with which the beginning and ending of transition phases was determined (dotted red and blue).
		\textbf{(D)} Probability density of dominance (blue) and transition durations (red), normalized to mean dominance period $\expect{T_\mathit{dom}}$.  Dominance periods below the $5\%$ quantile are defined as `short' (blue arrow) and dominance periods above the $95\%$ quantile as `long' (red arrow).
		\textbf{(E)} Comparison of the distribution of durations of `return' transitions, `forward' transitions, and `dominance' periods following long or short dominance periods (red and blue violin plots, respectively).  Medians were compared pairwise with a  two-sided Wilcoxon rank-sum test.  `*' indicates a p-value$<0.1$, `***' a p-value $< 10^{-3}$.
	}
	\label{fig:result_figure}
\end{figure}

\clearpage

\begin{figure}
	\centering
\includegraphics[width=0.7\linewidth]{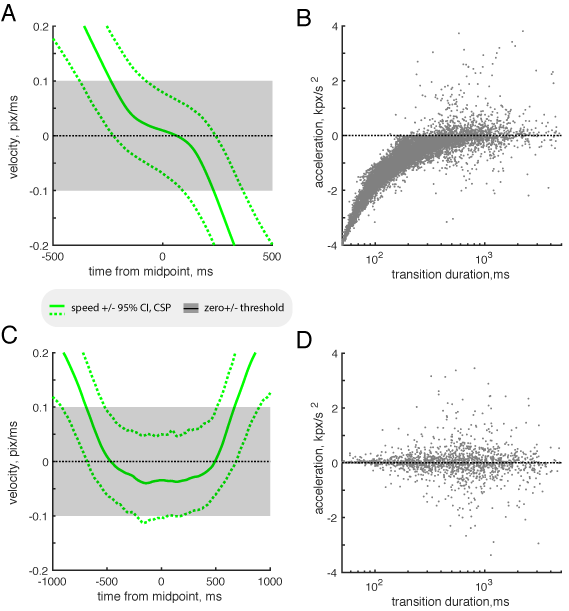}
	\caption[revision_results]{\figtitle{Velocity and acceleration during pursuit transitions.} 
       \textbf{(A)} Forward transitions of comparatively long duration (top quartile), average trajectory of velocity estimate (solid green) and confidence limits (dotted green), relative to midpoint of transition, and compared to low velocity thresholds $\pm0.1\mathit{pix/ms}$ (gray region).
		\textbf{(B)} All forward transitions, acceleration at midpoint (velocity difference between $-26\mathit{ms}$ before and $+26\%\mathit{ms}$ after midpoint), as a function of duration (in units of kilo-pixel, $\mathit{kpx}/\mathit{s}^2$).
       \textbf{(C)} Return transition of comparatively long duration (top quartile), average trajectory of velocity estimate (solid green) and confidence limits (dotted green), relative to midpoint of transition, and compared to low velocity thresholds (gray region).
		\textbf{(D)} All return transitions, acceleration at midpoint as a function of duration.
	}
	\label{fig:result_transition}
\end{figure}

\clearpage
\pagebreak
\section{Discussion}

We have further improved an established ``no-report'' paradigm for monitoring the phenomenal appearance of binocular rivalry display without soliciting volitional reports from the observers \citep{naber2011perceptual,frassle2014binocular,tsuchiya2015no}.  Our improved analysis -- `cumulative smooth pursuit' (CSP) -- has three main advantages: pursuit velocity is estimated continuously (not intermittently), pursuit phases are discriminated with a temporal resolution better than $\pm100\mathit{ms}$, and performance is robust for a wide range of oculomotor patterns.  Thus, our approach reveals additional details about the temporal progression of binocular rivalry and facilitates studies with developmental and patient cohorts.


Our results confirm and extend several conclusions from previous work \citep{naber2011perceptual,frassle2014binocular,kornmeier2012eeg}.  A reversal of direction in physical display motion is followed by a reversal in smooth pursuit direction (an `ocular response' in our terms) with an average latency of approximately $200\,\mathit{ms}$.  A manual motor response to the display reversal (`volitional report') follows after a further delay of approximately $250\,\mathit{ms}$.  Similarly, a reversal of apparent display motion is followed first by an `ocular response' and only approximately $250\,\mathit{ms}$ later by a `volitional report'.  An even larger lag  (approximately $350\,\mathit{ms}$) has been observed between electrophysiological correlates of perceptual reversals and manual responses to such reversals \citep{kornmeier2012eeg}, albeit for multi-stable displays other than binocular rivalry. 

A benefit of CSP is that pursuit velocity is estimated continuously (without intermittent gaps) and that confidence limits are provided at all times.  This contrasts to existing methods \citep{naber2011perceptual,frassle2014binocular}, which produce an intermittent estimate without confidence limits.   Where a `ground truth' may be established, CSP estimates were significantly less variable ($55\%$ smaller interquartile range of latencies) and more accurate ($15\%$ lower residual error), even under ideal conditions (highly linear pursuit episodes).  Overall, CSP estimates ranged over marginally higher values of velocity, acceleration, and jerk ($5\%$, $8\%$, and $25\%$, respectively), and thus appeared to be marginally more sensitive and/or marginally more volatile than existing methods. 

The reason for these differences is algorithmic.  Firstly, CSP filters eye position in an an adaptive (context-sensitive) manner ({\it i.e.}, by means of robust splining), whereas existing methods apply linear filtering, which is insensitive to context.   Secondly, CSP interpolates gaps in the eye position record such as to avoid discontinuities in acceleration (again by means of splining), whereas existing methods introduce such discontinuities (by interpolating linearly).  Thirdly, CSP estimates instantaneous velocity on the basis of an extended temporal neighbourhood ({\it i.e.}, by differentiating a robust spline), whereas existing methods compute velocity from a narrower basis ({\it i.e.}, from two successive position values).


The phenomenal appearance of rivalry displays comprises not only the two categorical alternatives ({\it i.e.}, leftward or rightward motion), but also intermittent or transitional appearances.  These transitional appearances are of considerable interest, in part because they are thought to reveal the respective contributions to rivalry dynamics of adaptation and noise \citep{brascamp2006time, pastukhov2011cumulative}.  In studies of binocular rivalry, observers are therefore often asked to distinguish three categories of phenomenal appearance: `leftward motion', `rightward motion', and `mixed/patchy' motion.  

Similar to phenomenal appearance, the velocity estimates of CSP exhibited a distinctly trimodal distribution (see Fig.\,\ref{fig:result_figure}A).  In addition to two modes for large rightward and leftward velocity, there was a third mode for near-zero velocities, suggesting that smooth pursuit did not merely alternate between rightward and leftward phases, but also lingered for extended periods at low velocities (see Fig.\,\ref{fig:result_transition}).  To quantify these phases of pursuit behaviour, we parsed CSP records into periods of `pursuit dominance' (rightward or leftward velocity) and periods of `pursuit transition' (low velocity).  The latter we  subdivided into `forward transitions' (leading from one dominance to another) and `return transitions' (leading back to the same dominance).   The temporal precision of this parsing was typically better than $\pm 100\,\mathit{ms}$ and reflected the local width of the confidence interval for pursuit velocity.

The succession of pursuit phases was fairly rapid, changing on average approximately once per second.   Specifically, periods of `pursuit dominance' lasted approximately two seconds, `return transitions' averaged approximately half a second and `forward transitions' approximately a quarter second.  Accordingly, it was not feasible to establish and compare the associated perceptual dynamics  from volitional reports.

Overall, the statistics of pursuit dominance phases resembled the results typically reported for perceptual dominance.   Pursuit dominance phases conformed to an approximately log-normal distribution and exhibited a weakly positive sequential dependence, as expected for perceptual dominance phases \citep{murata2003,vanee2009,pastukhov2011cumulative,cao2016}.

In an effort to compare {\it pursuit transitions} to their {\it perceptual} counterparts, we investigated the effects of perceptual adaptation, which is known to affect the latter \citep{pastukhov2011cumulative}.  Specifically, at times at which adaptation is weak, {\it perceptual} transitions take longer and are more likely to return to the previous dominance than at other times.  As a corollary, return transitions generally take longer than forward transitions.  The presumed reason is that transitions are thought to be driven partly by adaptation (drift) and partly by noise (diffusion) \citep{kim2006,brascamp2006,kang2010,pastukhov2011cumulative,pastukhov2013multi,arani2018changes}\footnote{When adaptation is weak, driving forces are absolutely smaller and have relatively larger stochastic component, resulting in longer transitions and more uncertain outcomes.}.

The results for {\it pursuit} transitions inferred from CSP records consistently mirrored results previously reported for {\it perceptual} transitions on the basis of volitional reports \citep{pastukhov2011cumulative}.  Firstly, `return transitions'  ($7.9\%\pm0.9$) were less frequent than `forward' transitions ($92.1\%\pm0.9$) and this proportion was almost identical to that reported previously for perceptual transitions ($9\%$ {\it versus} $91\%$) based on volitional reports. Secondly, `return' transitions took approximately twice as long as `forward' transitions, similar to the threefold difference observed previously for perceptual transitions.
Thirdly, both `forward' and `return' transitions took significantly {\it longer} at times at which perceptual adaptation was weak, again consistent with previously reported results for perceptual transitions.  These results are consistent with the possibility that oculomotor transitions closely reflect perceptual transitions. 

We conclude that `cumulative smooth pursuit' improves existing methods for monitoring binocular rivalry by means of recording optokinetic nystagmus \citep{naber2011perceptual,frassle2014binocular}.  By continuously estimating pursuit velocity within certain confidence limits, changes of oculomotor state may be detected with a precision better than $\pm100\mathit{ms}$.  How closely and faithfully this reflects the underlying dynamics of perceptual states remains to be determined, but the results about `forward' and `return' transitions, summarized above, are encouraging in this regard.   Being able to monitor binocular rivalry with higher temporal resolution may prove useful in several contexts, including the characteriziation of perceptual reversals, the validation of computational models of reversal dynamics, and the study of neurophysiological correlates of binocular rivalry.

\section{Source code repository}

The source code of the `Cumulative smooth pursuit' (CSP) analysis of eye position records is available free from the following repository:

\noindent
{\small
\url{https://github.com/cognitive-biology/Cumulative-smooth-pursuit-analysis-of-BR-OKN}
}

\section{Acknowledgment}

We acknowledge support from DFG BR987/3 to JB and from NKFI 110466 to IK, as well as helpful discussions with Alexander Pastukhov and P\'{e}ter Solt\'{e}sz.
We thank Zsolt Unoka (MD, PhD) and Kinga Farkas (MD, PhD), Department of Psychiatry and Psychotherapy, Semmelweis University, for recruiting and diagnosing patients.
Statement of author contributions: SA conceived, developed and performed CSP analysis; GZ designed, developed and performed behavioural experiments; all authors discussed the results and contributed to the final manuscript.


\end{document}